# Clean realization of the Hund physics near the Mott transition: NiS$_2$ under pressure

Ina Park[1], Bo Gyu Jang[2], Dong Wook Kim[1], Ji Hoon Shim[1,†], and Gabriel Kotliar[3,4]

[1]*Department of Chemistry, Pohang University of Science and Technology, Pohang 37673, Republic of Korea*

[2]*Department of Advanced Materials Engineering For Information & Electronics, Kyung Hee University, Yongin 17104, Republic of Korea*

[3]*Physics and Astronomy Department, Center for Materials Theory, Rutgers University, Piscataway, New Jersey 08854, USA*

[4]*Condensed Matter Physics and Materials Science Department, Brookhaven National Laboratory, Upton, New York 11973, USA*

[†]Email: jhshim@postech.ac.kr

**Abstract**:

Strong correlation effects caused by Hund's coupling have been actively studied during the past decade. Hund's metal, strongly correlated while far from the Mott insulating limit, was studied as a representative example. However, recently, it was revealed that a typical Mott system also exhibits a sign of Hund physics by investigating the kink structure in the spectral function of NiS$_{2-x}$Se$_x$. Therefore, to understand the Hund physics in a half-filled multi-orbital system near the metal-insulator transition, we studied pressure-induced metallic states of NiS$_2$ by using density functional theory plus dynamical mean-field theory. Hund physics, responsible for suppressing local spin fluctuation, gives low-energy effective correlations, separated from Mott physics, which suppresses charge fluctuation at higher energy. This effect is prominent when $J$ becomes comparable to the quasiparticle kinetic energy, showing apparent scaling behavior of the kink position $E_{\text{kink}} \sim J \cdot Z$. We suggest that the Hund effect can also be observed in the optical conductivity as a non-Drude-like tail with $1/\omega$ frequency dependence and non-monotonic temperature evolution of the integrated optical spectral weight at fixed frequency. Our study demonstrates the important role of Hund's coupling for electronic correlations even in a half-filled system.

The correlation effects from Hund physics have been actively studied recently in so-called Hund's metal, which contains various material examples such as iron pnictides, ruthenates, molybdates, and infinite-layer nickelates. [1-4] It successfully demonstrated the correlations due to the suppression of spin fluctuations, as it strongly renormalizes the quasiparticle bands but brings the system farther away from the Mott transition. [5,6] As a result, it mainly features high-spin local moment [7] and heavy effective mass despite the large quasiparticle bandwidth, where the latter is related to a kink structure inside the quasiparticle bands [2,3]. The quasiparticle bands are strongly renormalized in the vicinity of the Fermi level but soon become weakly correlated and incoherent, thus leaving a kink feature. This kink feature has been interpreted in terms of spin-orbital separation (SOS) of Kondo screening [8-12] or a resilient quasiparticle beyond the Fermi liquid (FL) energy [13,14], where the SOS stems from the nature of Hund's metal, neither non-singly-filled nor non-half-filled multi-orbital system.

Interestingly, however, a recent study [15] found that an archetypal Mott system $NiS_{2-x}Se_x$ [16-20] exhibits a similar kink structure inside the quasiparticle bands, surprisingly reminiscent of Hund's metals. In this study, the combined experimental and theoretical examination revealed that Hund's coupling $J$ plays a crucial role in correlated metallic states of $NiS_{2-x}Se_x$. It produces the kink structure and determines the characteristic quantities, such as quasiparticle coherence-incoherence crossover energy. Since the correlated metallic states of $NiS_{2-x}Se_x$ have been well understood by the Brinkman-Rice picture [21-23] and the orbital degree of freedom is already quenched due to its half-filled nature, this finding immediately raises a question of how Hund physics plays near the Mott metal-insulator transition (MIT) and further how they work together in a half-filled multi-orbital system.

The observed kink structure of $NiS_{2-x}Se_x$ implies that the correlation effects from $J$ are not merely additional to $U$ at high energy regions [6], but it also gives ununiform correlations at relatively low energy regions inside the quasiparticle bands. In this context, to investigate the Hund correlation effects in a half-filled multi-orbital system, we examined pressure-induced correlated metallic states of $NiS_2$ using density functional theory plus dynamical mean-field theory (DFT+DMFT). Compared to doping, hydrostatic pressure can homogeneously control the Ni $d$ bandwidth, which also offers a simpler parameter set of bonding distances, $d_{S\text{-}S}$ and $d_{Ni\text{-}S}$. [24-26]

In this article, first, we will verify that pressurized $NiS_2$ exhibits Hund correlation effects like Se-doped cases and investigate how $J$ generates the kink structure and low-energy effective correlations. The Hund vs. Mott physics in a half-filled multi-orbital system will be described by investigating the local impurity information and resulting spectral function. Then, we will study how large $J$ is needed to observe the regime with Hund correlations and propose that such a Hund regime could be defined with the kink position scaled by $J$, $E_{kink} \propto J \cdot Z$. Finally, we will suggest that Hund effects could also be manifested as a non-Drude-like optical conductivity and non-monotonic temperature dependence of optical spectral weight evolution, which could be observed experimentally.

## I. Kinks in correlated metallic states of $NiS_2$

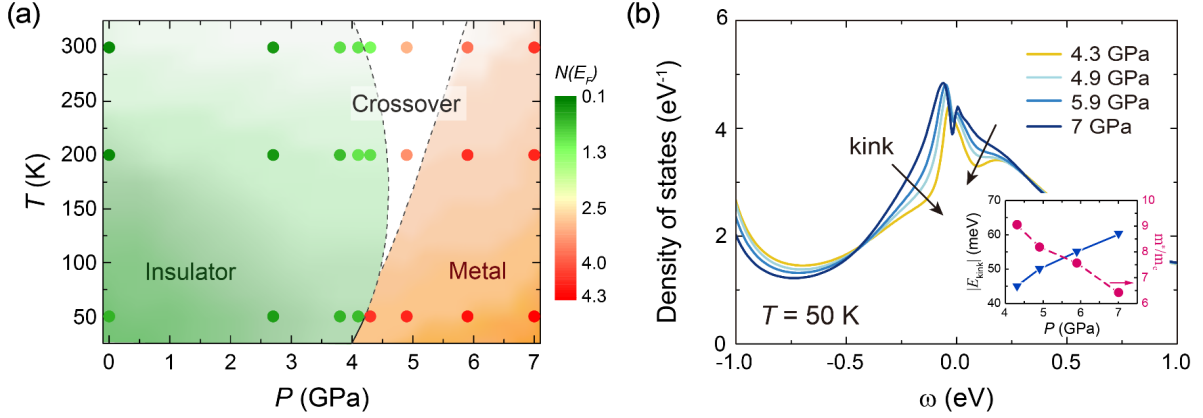

**Figure 1.** (a) *P-T* phase diagram of $NiS_2$ obtained from DFT+DMFT. The color of the dots represents the density of states value at the Fermi level $N(E_F)$, and the boundaries of the crossover region are marked with dashed lines. A solid line indicates an abrupt metal-insulator transition. (b) Quasiparticle density of states of metallic states at $T = 50$ K. Arrows indicate the kink evolution. (inset) Kink position on the hole side $|E_{kink}|$ (refer to the main text) (triangle) and effective mass $m^*/m_e$ (circle) as a function of pressure.

First, the overall *P-T* phase diagram obtained from the DFT+DMFT calculation is shown in Fig. 1(a). The crystal structures were optimized at $T = 50$ K for each pressure condition (see Methods), and the resulting phase diagram is qualitatively consistent with previously reported experimental results in critical pressure and temperature scales. [22] As shown in the change of $N(E_F)$, the density of states (DOS) value at the Fermi level, at $T = 50$ K, an abrupt phase transition occurs at around $P = 4$ GPa, as marked with the solid line. Notably, the bond lengths $d_{S-S}$ and $d_{Ni-S}$ showed an abrupt jump at the transition point, in accord with vibrational spectroscopy observations. [25,26] A continuous phase transition, named the crossover region, occurs at higher temperatures.

For metallic states at $T = 50$ K, one can see the kink structure inside the quasiparticle peak, as shown in Fig. 1(b). This kink structure is produced from the kink inside the self-energy, which will be discussed soon and can also be observed directly in the band structure (see Fig. S1). As pressure increases, as marked with arrows in Fig. 1(b), kink moves outward from the Fermi level, and at the same time, effective mass decreases, as those values are explicitly shown in the inset of Fig. 1(b). (Here, the kink position is defined as a frequency where it begins to deviate from the Fermi liquid behavior of the real part of the self-energy $Re[\Sigma(\omega)]$.) It can be understood that the FL behavior of the self-energy retains toward the higher frequency as it approaches the high-pressure FL regime.

## II. Hund correlation effects at a low-energy region

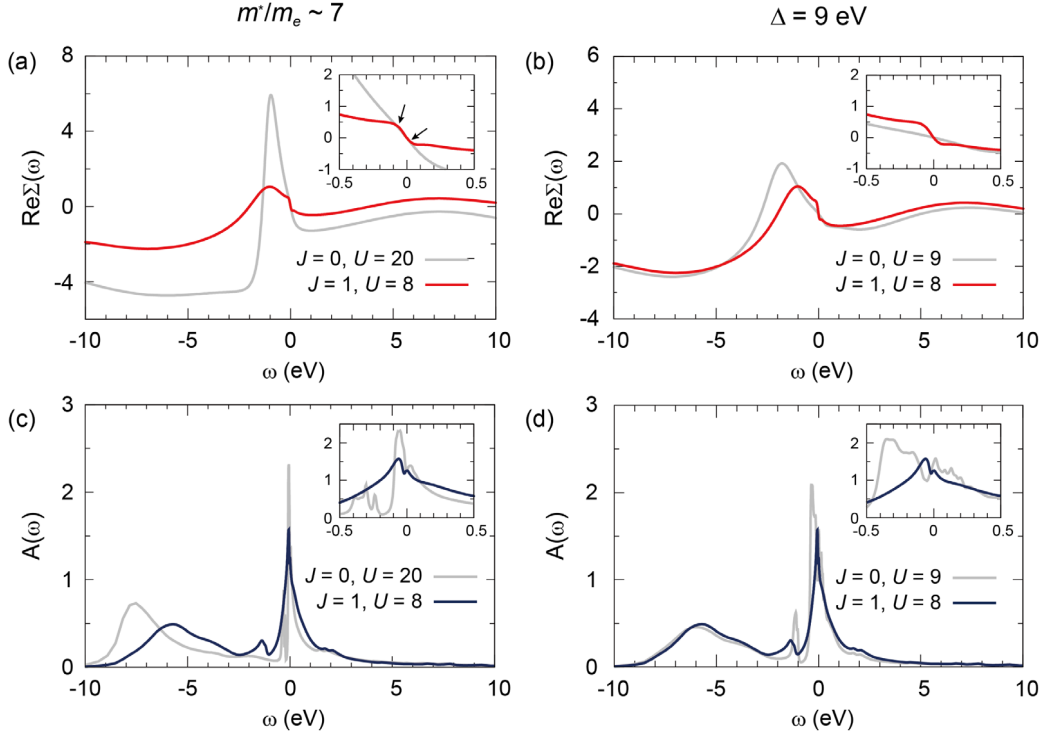

**Figure 2.** (a-b) Real part of the self-energy for $J = 0$ eV (grey) vs. $J = 1$ eV (red) when $m^*/m_e \sim 7$ (a) is fixed and $\Delta = 9$ eV (b) is fixed. For comparison between different parameter sets, the value of the self-energy at zero frequency is adjusted to zero. (c-d) Ni $e_g$ density of states for $J = 0$ eV (grey) vs. $J = 1$ eV (blue) when $m^*/m_e \sim 7$ (c) is fixed, and $\Delta = 9$ eV (d) is fixed. The pressure condition is $P = 5.9$ GPa, and insets show a low energy range near the Fermi level.

As mentioned, the kink inside the QP bands comes from the kink of the $\text{Re}[\Sigma(\omega)]$, marked with the arrows in the inset of Fig. 2(a). To investigate the effect of Hund's coupling $J$ on the kink structure and related low-energy correlations, first, we fixed the effective mass, *i.e.*, correlation strength, to about 7 and compared the $J = 0$ vs. 1 eV cases. The kink, produced by finite $J$, features a strong linear frequency dependence (FL behavior) only near the Fermi level and a sudden slope change followed by a weak frequency dependence. As a result, the quasiparticle spectral function $A(\omega)$ shows strongly renormalized peak only near the Fermi level and broad overall bandwidth, as shown in the inset of Fig. 2(c).

On the other hand, without $J$, a large value of $U = 20$ eV is required to produce the same effective mass, and it gives much more pronounced Hubbard bands, as shown in Fig. 2(c). Mainly, $\text{Re}[\Sigma(\omega)]$ is linear in frequency for much wider frequency range of around 1 eV (for the hole side), as well as the quasiparticle bands show coherent peak and strongly renormalized overall bandwidth. This is what is usually expected for the correlated metallic states in Mott(-Hubbard) physics, which has a renormalized quasiparticle band $Z\varepsilon_{\boldsymbol{k}}$, where $Z = m_e/m^*$ and $\varepsilon_{\boldsymbol{k}}$ is the bare band dispersion. However, the above comparison between the $J = 0$ vs. 1 eV cases shows that the correlation effects purely due to Hubbard $U$ should give a much higher Fermi liquid energy scale and coherent QP peak shape, which was not the experimental observation. [15]

Since both $U$ and $J$ intricately contribute to correlations near the MIT, next, we fixed the atomic charge transfer energy $\Delta = U + J = 9$ eV [5] and compared the $J = 0$ vs. 1 eV cases. By fixing $\Delta$, which represents the Mott physics – suppression of charge fluctuation, we can dispart the correlation effects at relatively high-energy regions and investigate the low-energy correlation effects due to $J$. As shown in Figs. 2(b) and 2(d), the self-energy and spectral function at high-energy regions are almost identical, as expected, while the low-energy region near the Fermi level shows significant differences. As magnified in the inset of Fig. 2(b), when $J$ is not considered, the self-energy shows FL behavior across a much wider range of frequencies. It is weakly correlated with a small effective mass of approximately 2, despite an equivalent amount of $\Delta$. DOS also reflects the similarities and distinctions. While the position of the Hubbard bands, its spectral weight, and overall quasiparticle bandwidth are almost identical, $J = 0$ eV also shows a weak renormalization, unlike the finite $J$ case.

As discussed in the introduction, the Brinkmann-Rice picture of renormalized quasiparticle bandwidth with quasiparticle residue Z has understood the correlated metallic state of Mott systems. However, the electronic structure of NiS$_2$ has a broad bandwidth despite the large effective mass, even though it is very near MIT. It implies that considering only Mott physics does not completely describe the correlated metallic states of NiS$_2$, which is a multi-orbital system. Hund physics also plays a key role.

## III. Mott vs. Hund - Atomic impurity information and spectral function

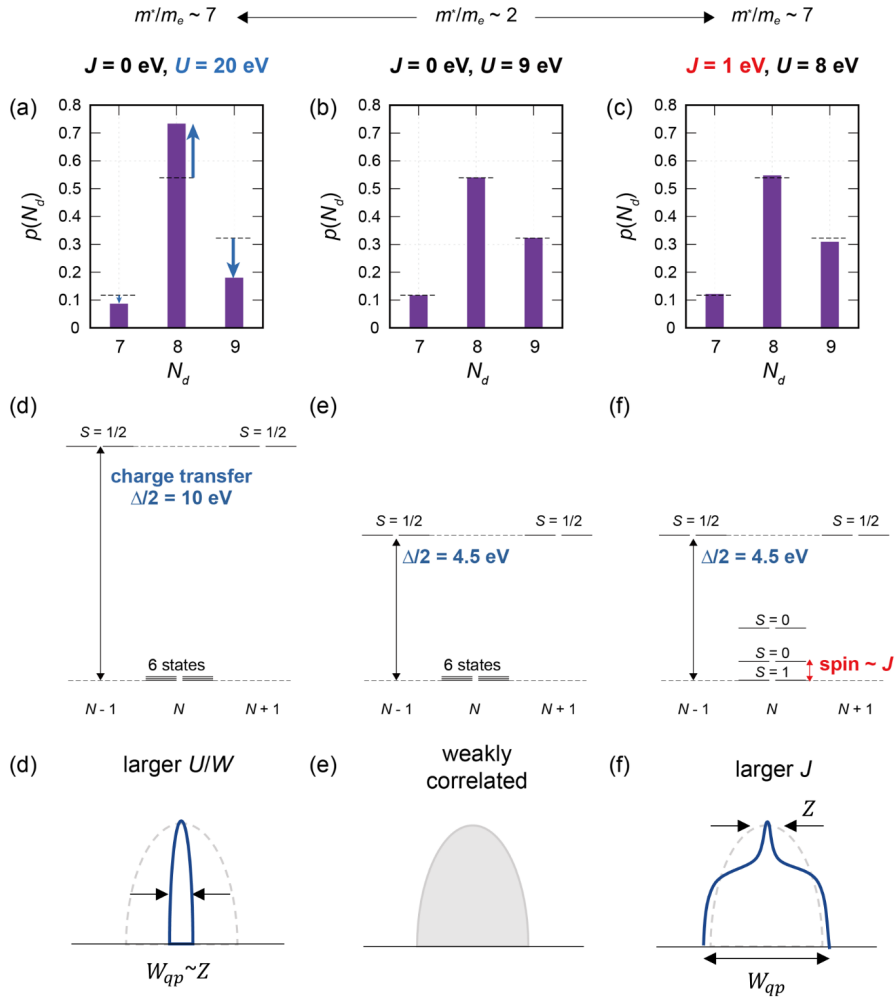

**Figure 3.** Comparison between the effective mass increase by increasing $U$ (left column) versus increasing $J$ with fixed $\Delta = U + J$ (right column) from the weakly correlated case (center column). (a-c) Impurity charge histogram for $N_d = 7$, 8, and 9 states. (d-f) Schematic impurity energy level diagram. (g-i) Schematic quasiparticle density of states.

Mott versus Hund physics in a half-filled multi-orbital system, which affects the relatively high and low-frequency regions, respectively, can be more clearly understood by looking into the atomic impurity information of $NiS_2$. In Fig. 3, we compared the two routes of increasing effective mass − 1) by increasing Hubbard $U$ (or $\Delta$) and 2) by increasing Hund's coupling $J$. The weakly correlated state with an effective mass of around 2 is the starting point, obtained from parameters $J = 0$ eV and $U = 9$ eV (*i.e.*, $\Delta = 9$ eV). As shown in Fig. 3(b), it has the probabilities of $N$-1 ($N_d$=7) and $N$+1 ($N_d$=9) occupancy states as comparable to $N$ ($N_d$=8) states. (Here, $N_d$ is the number of electrons in the Ni $d$ orbital.) The charge fluctuation is not suppressed much in this weakly correlated state.

Then, if we take the first route of increasing $U$, the probabilities of $N$, $N$-1, and $N$+1 states significantly change. $N$ state now becomes greatly dominant with a probability of about 75%, and this suppression of charge fluctuation increases the effective mass to about 7. This route stands for the conventional Mott physics. On the other hand, if we take the second route of increasing $J$, the probabilities of $N$-1, $N$, and $N$+1 states remain almost identical even though the effective mass increases the same amount. These invariable probabilities are straightforward since $\Delta$ is not changed for the second route. Therefore, it stands for the low-energy effective correlations from Hund physics, where the increase of the effective mass is solely due to the role of $J$ generating spin multiplet structure, which will be discussed soon.

In Figs. 3(d)-3(f), we depicted the schematic energy diagram of $N$-1, $N$, $N$+1 states for each parameter sets. (The excited states of the $N$-1 state are omitted for the sake of simplicity.) Starting from the weakly correlated states, shown in Fig. 3(e), where the charge excitation requires the energy cost of $\Delta/2 = 4.5$ eV and all 6 $N$ states are degenerate due to zero $J$, the charge excitation energy increases up to 10 eV by increasing only $U$, as shown in Fig. 3(d). As a result, the probabilities of $N$-1 and $N$+1 states are greatly decreased.

On the other hand, if we increase $J$, as shown in Fig. 3(f), the charge excitation energy does not change, and now $N$ states have a spin multiplet structure due to finite $J$. $S = 1$ high-spin states become the ground states and interorbital and intraorbital $S = 0$ low-spin states becomes the first and the second excited states, respectively. The decreased ground state degeneracy results in the effective mass increase, which can be understood as due to the reduced number of effective Kondo coupling channels. [5,27] (The resulting spin-resolved atomic configuration histogram is shown in Fig. S2.)

The two routes discussed in Fig. 3 can then be viewed as two governing physics of strong electronic correlations in a half-filled multi-orbital system − 1) Mott physics responsible for charge blocking and 2) Hund physics for spin blocking. These two routes give clear traces in the QP DOS as we investigated. As schematically described in Figs. 3(g)-3(i), Mott physics gives narrow overall renormalized QP bandwidth with reduced spectral weight, while Hund physics gives a renormalized peak at the center but with broad overall bandwidth, characterized by a kink structure.

**IV. Hund regime of the half-filled multi-orbital system**

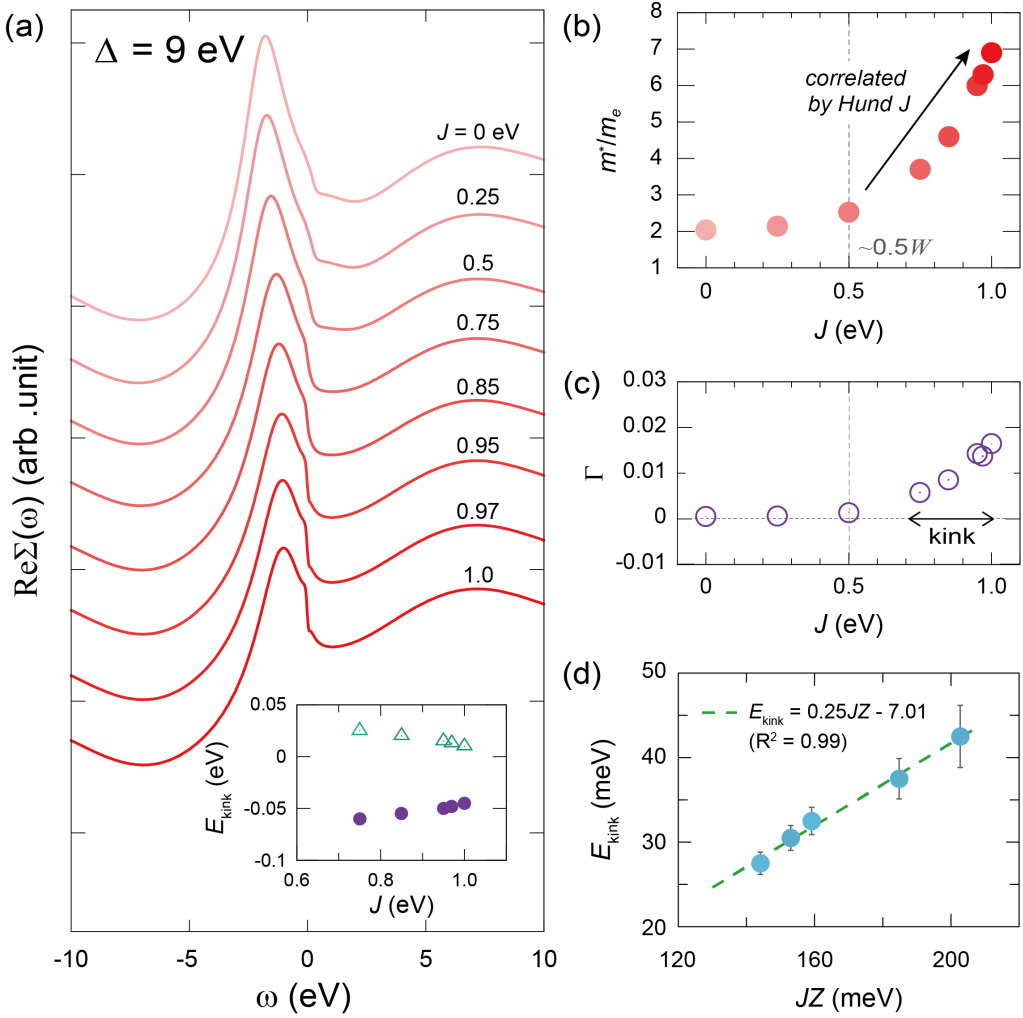

**Figure 4.** (a) Real part of the self-energy with various values of $J$ from 0 (top) to 1 eV (bottom) with fixed $\Delta = 9$ eV. (inset) Kink positions as a function of $J$. Green dots are for electron side kinks, and purple dots are for hole side kinks. (b) Effective mass as a function of $J$. Dashed line marks the energy equivalent to $0.5 \cdot W$, where $W$ is the half bare bandwidth. (c) Quasiparticle lifetime $\Gamma$, defined as $\Gamma = -Z\text{Im}\Sigma(i0^+)$ as a function of $J$. The region of $J$ with an apparent kink feature in the self-energy is marked with a double-sided arrow. (d) Scaling relation between the average kink position $E_{\text{kink}}$ and $J \cdot Z$ in the region of $J$ with apparent kink feature. The error bars indicate how prominent the kink feature is (see main text.)

Then, we investigated how large $J$ is needed to generate an apparent Hund correlation effect. Here, we gradually increased $J$ from 0 to 1 eV with fixed $\Delta = 9$ eV, and the resulting self-energies are plotted in Fig. 4(a). One can see that the kink structure becomes distinct, and the slope at the zero frequency changes abruptly as $J$ becomes roughly larger than 0.5 eV. This change is more explicitly shown if we see the effective mass as plotted in Fig. 4(b). At $J = 0$ eV, the effective mass is approximately 2 and barely increases until $J$ approaches 0.5 eV. However, once $J$ becomes greater than 0.5 eV, the effective mass experiences a drastic rise with increasing $J$. This region with a large effective mass produced by $J$ can be interpreted as a 'Hund regime,' where the distinguished Hund correlation effect is exhibited in a half-filled multi-orbital system.

In this Hund regime, as shown in the inset of Fig. 4(a), the average kink position $E_{\text{kink}}$, $(E_{\text{kink}}^{\text{hole}} + E_{\text{kink}}^{\text{el}})/2$, shows a clear decreasing trend with increasing $J$. The quasiparticle scattering rate $\Gamma$ also significantly increases with increasing $J$ in this region, as shown in Fig. 4(c). It is consistent with the previous report that the kink energy is intimately related to the quasiparticle coherence-incoherence crossover energy. [15] If we compare this criterion of $J \sim 0.5$ eV to the half bare bandwidth, $W$, we can expect the system to enter the Hund regime when $J$ becomes comparable to $W$, *i.e.*, $J \sim 0.5W$. (Here, 'bare' means 'without Hund correlation,' so it is the half bandwidth obtained from parameters $U = 9$ eV and $J = 0$ eV. Notably, the results do not change qualitatively even if we use DFT bandwidth instead.) In the case of $NiS_2$, $J = 1$ eV and $W \sim 1$ eV, so the Hund effect is compelling in this system.

In the Hund regime, we found that the kink energy is scaled by the relation $E_{kink} \sim J \cdot Z$, as shown in Fig. 4(d). This scaling relation is consistent with the observation that $J$ decreases both $E_{kink}$ and $Z$ and holds for the case of pressure-induced kink evolution as well. As shown in Fig. S3, we observed the scaling relation with a similar coefficient of 0.38 as we increased pressure. In this case, we change $Z$ instead of $J$ by changing $W$. In Fig. 4(d), the error bar is calculated from the slope difference of $Re\Sigma(\omega)$ before and after the kink (refer to supplementary section IV.), and it indicates how prominent the kink feature is, *i.e.*, how it is close to the Hund regime. The increasing size of the bar implies the system has a weaker Hund effect.

Even though the underlying physical origin of this scaling relation is not fully understood, we expect it would be related to Kondo physics. Without $J$, it was studied that the kink can manifest at the Kondo energy scale inside the quasiparticle bands of the correlated metallic state. [28,29] For a half-filled multi-orbital system with finite $J$, however, the impurity local moment can also fluctuate between its own high- and low-spin states. The local moment fluctuation controlled by $J$ also comes into play and may give the $J$-dependent coefficient to the kink position scaling.

## V. Hund effect in optical conductivity

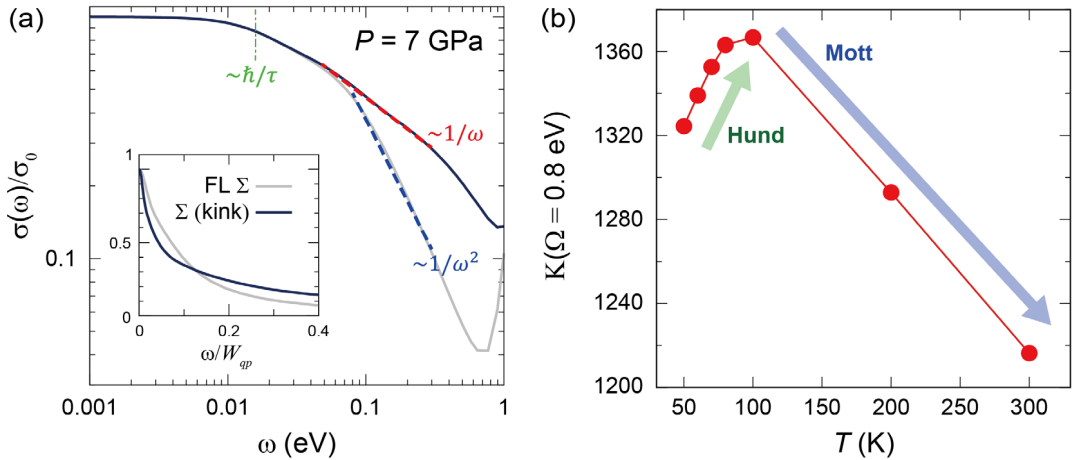

**Figure 5.** (a) Optical conductivity normalized to its value at zero frequency, $\sigma(\omega)/\sigma_0$, obtained from the original self-energy with kink structure (blue) and the Fermi liquid self-energy (grey) at $P = 7$ GPa. (inset) Normalized optical conductivity as a function of $\omega/W_{qp}$. (b) Optical spectral weight $K(\Omega)$ at $\Omega = 0.8$ eV as a function of temperature at $P = 7$ GPa.

Finally, we investigated how the trace of the Hund effect manifests in optical conductivity. Since the Hund effect redistributes spectral weight, not renormalizing the overall quasiparticle bandwidth, it can give a non-Drude-like tail of optical conductivity. In Fig. 5(a), the normalized optical conductivity $\sigma(\omega)/\sigma_0$ obtained from the original self-energy with kink and the conductivity from the Fermi liquid self-energy (see Fig .S4) [30] are shown for pressure condition *P* = 7 GPa. Both conductivities start to decrease at a frequency near the quasiparticle scattering rate $\hbar/\tau$. Still, the frequency dependence at the intermediate frequency region, $\hbar/\tau < \omega < W_{qp}$, is significantly different, where $W_{qp}$ is the quasiparticle half bandwidth. The original conductivity shows non-Drude-like $\sigma(\omega) \propto 1/\omega$ behavior unlike the FL conductivity showing $\sigma(\omega) \propto 1/\omega^2$ behavior. If we compare these two as a function of $\omega/W_{qp}$, this results in a lagging tail of the original conductivity, as shown in the inset.

This tail feature gives a surplus optical spectral weight compared to the FL conductivity, as shown in Fig. S4, and it shows a specific temperature evolution, representing the Hund effect. In Fig. 5(b), the integrated optical spectral weight $K(\Omega) = \int_0^\Omega \sigma(\omega)$ at fixed frequency $\Omega$ = 0.8 eV, for example, is shown as a function of temperature. (Raw optical conductivity and $K(\Omega)$ for the whole frequency range is shown in Fig. S5.) At relatively low-temperature region, $K(\Omega = 0.8 \text{ eV})$ increases with decreasing temperature, which is reminiscent of iron pnictides [31]. This increasing trend of $K(\Omega)$ is peculiar since a typical Mott behavior should show a decrease in $K(\Omega)$ with increasing temperature. As a phase diagram describes, the Mott behavior of $NiS_2$ is recovered at temperatures above 100 K, as a positive $T_{MIT}$ slope implies.

Therefore, the Hund effect can be observed at low enough temperatures in half-filled multi-orbital systems. Note that the Hund regime would be very broad for the pressure axis. For example, the bare bandwidth of $NiS_2$ changes only about 10% as we increase pressure to 15 GPa, so *J/W* is still close to 1. As a result, the scaling relation of kink energy holds toward 15 GPa. (See Fig. S3)

In this article, we demonstrated that a typical Mott system $NiS_2$ exhibits apparent Hund correlation effects near MIT, giving traces in spectral function and optical conductivity. Mott and Hund physics are dominant in this system, and we examined the high- and low-energy correlations originating from charge and spin blocking, respectively. We proposed that the Hund regime can be defined with the scaling relation $E_{kink} \sim J \cdot Z$ and the proportional relation between the optical spectral weight at fixed frequency and temperature.

## Acknowledgments

We greatly thank A. Georges, CJ Kang, and C. Kim for the fruitful discussion and comments. G. K. was supported by NSF-DMR1733071.

# Supplemental Material For

# Clean realization of the Hund physics near the Mott transition: NiS$_2$ under pressure

Ina Park[1], Bo Gyu Jang[2], Dong Wook Kim[1], Ji Hoon Shim[1,†], and Gabriel Kotliar[3,4]

[1]*Department of Chemistry, Pohang University of Science and Technology, Pohang 37673, Republic of Korea*
[2]*Department of Advanced Materials Engineering For Information & Electronics, Kyung Hee University, Yongin 17104, Republic of Korea*
[3]*Physics and Astronomy Department, Center for Materials Theory, Rutgers University, Piscataway, New Jersey 08854, USA*
[4]*Condensed Matter Physics and Materials Science Department, Brookhaven National Laboratory, Upton, New York 11973, USA*

[†]Corresponding author. Email: jhshim@postech.ac.kr

**Table of Contents**

## I. Computational Methods

In this paper, we investigated the pressurized states of NiS$_2$, and the effect of hydrostatic pressure is applied by the isotropic change of lattice constants. Most of the lattice constants are from experimental values [24], and some were linearly interpolated from the experimental lattice constants.

At every pressure condition considered in this work, the internal atomic positions are optimized while retaining the cubic symmetry at the DFT+DMFT level, where the DFT+DMFT calculation was performed as implemented in the DFT + embedded DMFT Functional (eDMFT) code. [32] The DFT calculation with full-potential augmented plane wave method was performed by using WIEN2k code [33], where the Perdew-Burke-Ernzerhof (PBE) generalized gradient approximation (GGA) was used for the exchange-correlation functional. [34] A 12x12x12 k-point mesh was used for the electronic self-consistent calculation.

Embedded DMFT considers the electronic correlation effect of Ni 3$d$ orbitals on top of an effective one-electron Hamiltonian generated by the DFT calculation. In the DMFT loop, the real harmonics basis was used, and the local axis rotation was also considered for the local basis of impurity atoms to reflect the local octahedral ligand field environment generated by sulfur atoms. All DMFT calculations are charge self-consistent calculations, and we used the density-density form of Coulomb interaction for interaction Hamiltonian with Slater parametrization. The hybridization energy window from -10 to 10 eV was generally considered to correctly describe the highly pressurized states with fixed corresponding band ranges. The parameters $U$ = 8 eV and $J$ = 1 eV were used, which describe the experimental $P$-$T$ phase diagram qualitatively well. If a different value of $J$ was used for comparison, it is specified for each case in the main text. The impurity model was solved by a continuous-time quantum Monte Carlo (CTQMC) impurity solver. A 17x17x17 k-point mesh and a frequency window from -10 to 10 eV are used to calculate optical conductivity.

In eDMFT code [32], Slater parametrization is used to consider Coulombic interaction. For correlated $d$-electrons, $F^0 \equiv U_{params}, F^2 \equiv 112/13\, J_{params}$, and $F^4 \equiv 70/13\, J_{params}$ are used for the convention, where $U_{params}$ and $J_{params}$ are $U$ and $J$ mentioned in the main text, respectively. If only $e_g$ orbitals are correlated, it can be equivalent to the Kanamori parametrization so that the excitation energy $\Delta E_a$ in the main text becomes equivalent to $J_K^{e_g}$.

## II. Kink structure inside the momentum-resolved spectral function of the correlated metallic state

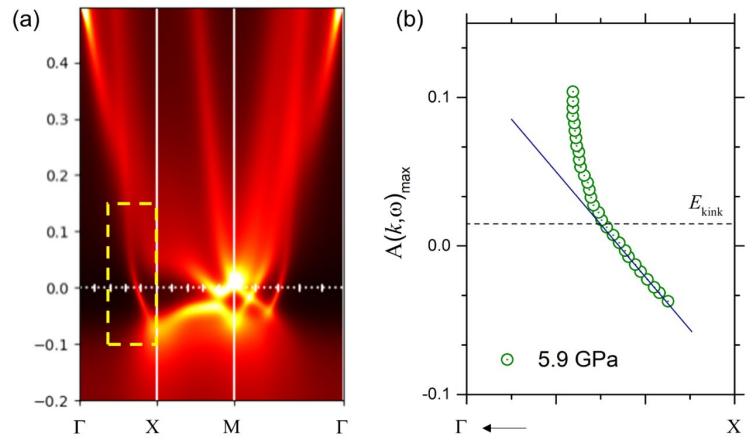

**Figure S1.** (a) Momentum-resolved spectral function A($k,\omega$) of NiS$_2$ at $P$ = 5.9 GPa and $T$ = 50 K. (b) Maximum weight point of the momentum-resolved spectral function, A($k,\omega$)$_{max}$, of the band crossing the Fermi level along Γ-X path. The momentum range is marked with the yellow dashed box in (a). The dashed horizontal line indicates the $E_{kink}$, the kink energy obtained from the real part of the self-energy. The blue solid is the guideline for the eye.

## III. Mott vs. Hund physics in the impurity histogram of $N$ states

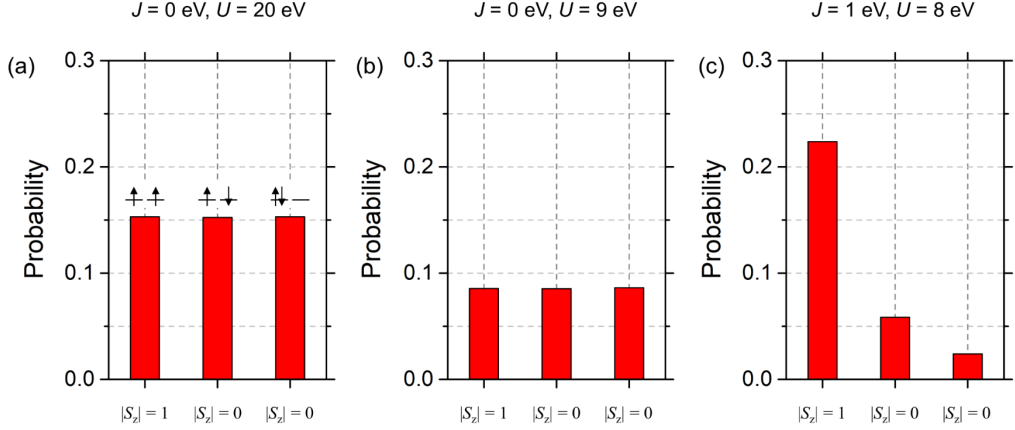

**Figure S2.** Impurity histogram of $e_g$ states for $N_d = 8$ occupancy obtained from parameters $J = 0$ eV, $U = 20$ eV (a), $J = 0$ eV, $U = 9$ eV (b), and $J = 1$ eV, $U = 8$ eV (c).

For $N(N_d = 8)$ charge states, since the $e_g$ configurations are dominant configurations in NiS$_2$ and for the sake of simplicity, we displayed only the histogram for six possible configurations of half-filled $e_g$ orbitals – high-spin $S = 1$ states ($|\uparrow, \uparrow\rangle$ and $|\downarrow, \downarrow\rangle$), inter-orbital low-spin $S = 0$ states ($|\uparrow, \downarrow\rangle$ and $|\downarrow, \uparrow\rangle$), and intra-orbital low-spin $S = 0$ states ($|\uparrow\downarrow, 0\rangle$ and $|0, \uparrow\downarrow\rangle$).

The center column is according to the weakly correlated state with an effective mass of around 2. Since the value of $J$ is zero, all the high- and low-spin states have the same probability due to their degeneracy. If we increase $U$, all three values are significantly increased but still equivalent. On the other hand, if we increase $J$ instead, the ground state degeneracy is broken, favoring the high-spin state. As a result, the high-spin state becomes dominant while the inter-orbital and intra-orbital low-spin states have significantly decreased probabilities.

## IV. Kink energy scaling relation for the case of pressure-induced kink evolution

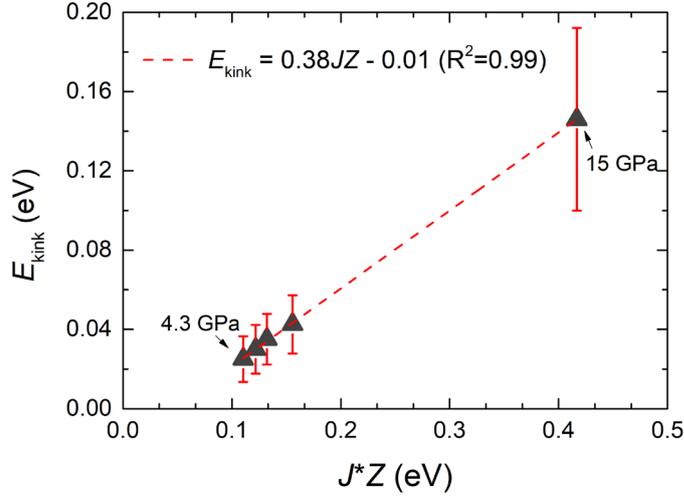

**Figure S3.** The Kink energy scaling relation for the pressure range from $P$ = 4.3 GPa to 15 GPa at $T$ = 50 K. Dashed line results from linear fitting and error bar indicating how much kink is pronounced in the self-energy. (Refer to the text below.)

    As pressure increases, the kink moves outward to the Fermi level, as shown in Fig. S3. We found that the same scaling relation proposed in the main text, $E_{kink} \sim J \cdot Z$, holds very well in this case, too. For pressure-induced kink evolution, it can be thought of as we change $Z$ by changing $W$, while $J$ is fixed to 1 eV. The proportional constant of this scaling relation was obtained as 0.38, which is not very different from the case of $J$ evolution, 0.22.

    In this figure, the error bars indicate how much the kink structure is pronounced. In this paper, we focused on one property of the kink structure of $NiS_2$, the abrupt slope change of the real part of the self-energy *followed by an almost frequency-independent part*. For example, at $P$ = 7 GPa, the slope of the Re$\Sigma(\omega)$ at zero frequency is -5.4, but it suddenly decreases to -0.6 after the kink. (If we look at the hole side.) So, here, we tried to measure 'how much kink is pronounced' by measuring 'how much slope changes.' The exact equation for the error bar is then:

$$c/\text{abs}(\log(\lambda_l - \lambda_{FL}))$$

where $\lambda_l$ and $\lambda_{FL}$ is the slope of the Re$\Sigma(\omega)$ for linear frequency dependence before (left kink) the kink frequency and the slope at the zero frequency (FL slope), respectively. (Here, the constant $c$ does not have a quantitative meaning, and the value of 0.01 was used to show the trend of error size clearly.) As pressure increases, the size of the error bar increases. It implies that the kink structure becomes weaker as pressure increases as it approaches the Fermi liquid limit. Therefore, even though the 15 GPa data follows the scaling relation, we may interpret it as the pressure condition is quite close to the crossover toward the FL state. For the $J$ evolution case shown in Fig. 4 in the main text, the value of 0.006 was used for $c$.

# V. Surplus optical spectral weight from kink structure

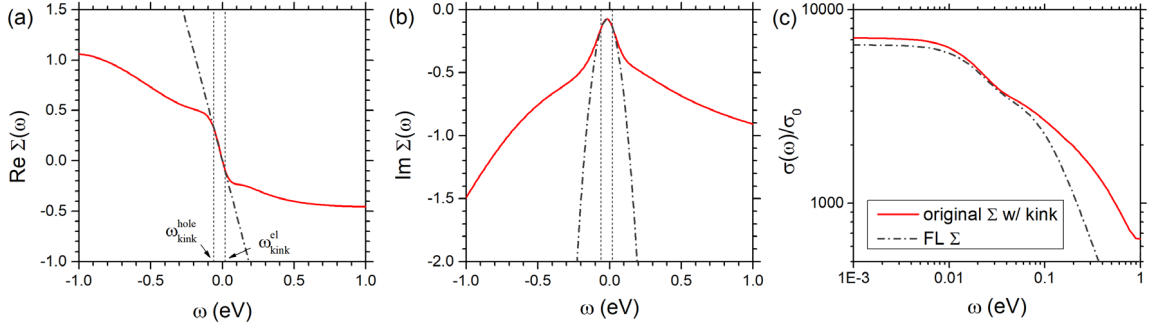

**Figure S4.** Real (a) and Imaginary (b) part of the self-energy with original kink structure (red solid) and Fermi liquid behavior (grey dash-dot). Dotted lines mark the kink position on the hole and electron side, $\omega_{kink}^{hole}$ and $\omega_{kink}^{el}$. (c) Optical conductivity obtained from the original self-energy (red solid) and FL self-energy (grey dash-dot). The pressure condition is $P$ = 7 GPa.

Here, the Fermi-liquid self-energy, FL $\Sigma$, is obtaind from fitting the self-energy with kink structure obtained from the charge-self-consistent DFT+DMFT calculation (original $\Sigma$) to the Fermi liquid (FL) form:

$$\Sigma(\omega, T) = \left(1 - \frac{1}{Z}\right)\omega - i\Omega_0^{-1}\left(\omega^2 + \frac{1}{2\Omega_0^{-1}Z}\Gamma\right).$$

, which has a linear and quadratic frequency dependence for real and imaginary parts, respectively. Here, $\Gamma$ is the quasiparticle scattering rate, $\hbar/\tau_{qp}$ [35], which can be directly obtained from the value of the imaginary part of the self-energy at zero frequency at a certain temperature condition. For example, the original and FL self-energy of $NiS_2$ for pressure condition of $P$ = 7 GPa is shown in Fig. S4(a) and S4(b). The self-energy follows the FL behavior only near the Fermi level, where the frequency is smaller than the kink energy, $\omega_{kink}^{hole} < \omega < \omega_{kink}^{el}$, as marked with dotted lines. Very weakly frequency-dependent self-energy is followed by the kink.

Fig. S4(c) shows the optical conductivity obtained from the original $\Sigma$ and FL $\Sigma$. Near zero frequency, the two are almost identical, but as the frequency becomes the kink energy, the original conductivity starts to deviate from FL conductivity significantly, showing the surplus optical spectral weight inside the QP frequency range. Unlike the FL self-energy, which substantially reduces QP spectral weight, the original self-energy pushes away the spectral weight extremely weakly due to the substantial deviation from the linear dependence.

## VI. Temperature evolution of the optical conductivity and its integrated spectral weight

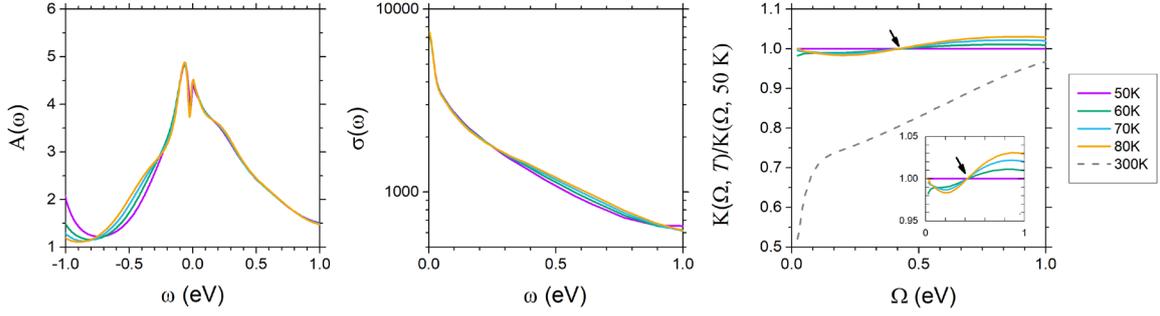

**Figure S5.** Quasiparticle spectral function A(ω) (a), optical conductivity (b) for temperature range $T$ = 50-80 K. (c) Integrated optical spectral weight normalized to it at 50 K, *i.e.*, K(Ω, $T$)/K(Ω, $T$ = 50 K) for temperature range $T$ = 50 – 300 K. Arrows mark the frequency value at which the integrated optical spectral weight crosses. Inset magnifies the temperature range from 50 K to 80K.

As discussed in the main text, the surplus optical spectral weight inside the quasiparticle frequency range is due to Hund physics in NiS$_2$. The Hund effect, generating spin multiplet structure inside the *N* states, increases the quasiparticle's effective mass but does not change the overall quasiparticle spectral weight. This phenomenon can be observed in the integrated optical spectral weight, too. Even though the temperature range for this phenomenon to be clearly manifested is limited to relatively low temperatures because the Mott physics will eventually dominate at relatively high temperatures, here, we can clearly see the redistribution of optical spectral weight inside the quasiparticle frequency range.

In Fig. S5(a) and S5(b), we can see the spectral weight or optical conductivity is slightly reduced as temperature increases at a relatively low-frequency range, $-0.3 \lesssim \omega \lesssim 0.2$, and then increased at relatively high-frequency range, $\omega \lesssim 1.0$. (We define the quasiparticle half bandwidth as 1 eV roughly.) This results in the integrated optical spectral weight K(Ω) crossing point, as marked with arrows in Fig. S5(c). This crossing means that the loss of spectral weight due to renormalization at the low-frequency range near the Fermi level is recovered around this crossing point, representing the Hund correlation effects. For the high-temperature condition where Mott physics becomes dominant, as shown with a grey dashed line in Fig. S5(c), the overall QP spectral weight should decrease for all frequency ranges since the spectral weight should escape to Hubbard bands.